\documentclass[conference, 10pt]{IEEEtran}

\usepackage{graphicx}
\usepackage{color}
\usepackage{placeins}
\usepackage{float}
\usepackage{tabularx,colortbl}

\usepackage{amssymb}
\usepackage{upgreek}

\usepackage{amsfonts,amsmath,amssymb}
\usepackage[usenames, dvipsnames]{xcolor}
\usepackage[labelfont=bf]{caption}

\providecommand{\norm}[1]{\left \lVert#1 \right  \rVert}

\newcommand{\from}{{\ensuremath{\colon}}}           

\providecommand{\mc}[1]{\mathcal{#1}}

\newcommand{\Real}{\mathbb{R}}

\usepackage[colorlinks=true,linkcolor=magenta]{hyperref}
\usepackage[sort&compress,comma,square,numbers]{natbib}
\usepackage{url}

\newcommand{\mmmr}[1]{M$^3$R}

\newcommand{\bxi}{\mathbf{\xi}}

\providecommand{\mc}[1]{\mathcal{#1}}

\begin{document}

\title{MIGRAINE:  MRI Graph Reliability Analysis and Inference for Connectomics}

\author{\IEEEauthorblockN{William Gray Roncal\IEEEauthorrefmark{1}\IEEEauthorrefmark{2},
Zachary H. Koterba\IEEEauthorrefmark{1},
Disa Mhembere\IEEEauthorrefmark{2},\\
Dean M. Kleissas\IEEEauthorrefmark{1} 
Joshua T. Vogelstein\IEEEauthorrefmark{3}\IEEEauthorrefmark{2}\IEEEauthorrefmark{4}
Randal Burns\IEEEauthorrefmark{2} 
Anita R. Bowles\IEEEauthorrefmark{5} \\
Dimitrios K. Donavos\IEEEauthorrefmark{5} 
Sephira Ryman\IEEEauthorrefmark{6} 
Rex E. Jung\IEEEauthorrefmark{6} 
Lei Wu\IEEEauthorrefmark{6} \IEEEauthorrefmark{7} 
Vince Calhoun\IEEEauthorrefmark{6} \IEEEauthorrefmark{7} 
and
R. Jacob Vogelstein\IEEEauthorrefmark{1}\IEEEauthorrefmark{2}}

\IEEEauthorblockA{\IEEEauthorrefmark{1}JHU Applied Physics Laboratory, Laurel Maryland 20723, USA.  Email: willgray@jhu.edu}
\IEEEauthorblockA{\IEEEauthorrefmark{2}Johns Hopkins University, 3400 N Charles Street, Baltimore, Maryland 21218, USA}
\IEEEauthorblockA{\IEEEauthorrefmark{3}Duke University, Durham, NC 27708, USA}
\IEEEauthorblockA{\IEEEauthorrefmark{4}Child Mind Institute, 445 Park Avenue, New York, NY 10022, USA}
\IEEEauthorblockA{\IEEEauthorrefmark{5}University of Maryland, Center for Advanced Study of Language, 7005 52nd Avenue, College Park, MD 20742, USA}
\IEEEauthorblockA{\IEEEauthorrefmark{6}University of New Mexico, 1 University Blvd NE, Albuquerque, NM 87131, USA}
\IEEEauthorblockA{\IEEEauthorrefmark{7}The Mind Research Network, 1101 Yale Blvd NE, Albuquerque, NM 87106, USA}
}

\maketitle

\begin{abstract}
Currently, connectomes (e.g., functional or structural brain graphs) can be estimated in humans at   $\approx 1~mm^3$ scale using a combination of diffusion weighted magnetic resonance imaging, functional magnetic resonance imaging and structural magnetic resonance imaging  scans. This manuscript summarizes a novel, scalable implementation of open-source algorithms to rapidly estimate magnetic resonance connectomes, using both anatomical regions of interest (ROIs) and voxel-size vertices. To assess the reliability of our pipeline, we develop a novel nonparametric non-Euclidean reliability metric.  
Here we provide an overview of the methods used, demonstrate our implementation, and discuss available user extensions. We conclude with results showing the efficacy and reliability of the pipeline over previous state-of-the-art.
\end{abstract}

\IEEEpeerreviewmaketitle

\section{Introduction}

The ability to estimate a connectome, i.e. a description of connectivity in the brain of an individual, promises advances in many areas from personalized medicine to learning and education, and even to intelligence analysis \cite{Sporns2010,Lichtman2008}. A robust analysis of these brain-graphs is on the horizon due to recent efforts to collect large amounts of multimodal magnetic resonance (\mmmr~) imaging data \cite{VanEssen2012a,Mennes2012}. An ideal methodology would enable scalable computing of graphs and functionals thereof in a way that yields estimates that are reliable. Moreover, such a tool would be open source, and make the data it processes open source in a user friendly way.  

Building such a tool, however, is challenging.  The data for each subject consists of about 1 gigabyte (GB) of \mmmr~ data.  Converting from raw data to graphs and functions thereof requires daisy-chaining over 20 subroutines, each of which implements a different transformation of the data. 
MRCAP is an existing pipeline that we previously developed for inference of graphs \cite{Gray2011}.  However, MRCAP has robustness and scalability limitations;  it requires about 10 hours per subject to generate a final output, and has scheduler constraints.  Moreover, MRCAP only generates small graphs, with 70 vertices, rather than voxel-wise graphs and functions thereof. Other pipelines (for example, \cite{Cui2013}) have similar problems.

The utility of the output brain-graphs is a function of its scientific meaningfulness.  Because such data currently lack ground truth for the estimates, or even other gold standards, investigators are left to assess the quality of estimates using only the data itself.  Test-retest datasets consist of multiple subjects, each of whom have been scanned multiple times. Most previous work on reliability has assessed parametric reliability of features of the data.  For example, it is standard to compare the between and within variances of, say, the number of edges \cite{Braun2012}.  However, these approaches are limited because they make parametric assumptions, and test scalar functions of the graphs.

We present here an MRI Graph Reliability Analysis and INference for ConnEctomics (MIGRAINE). methodology and associated software package.  In addition to satisfying the above two mentioned desiderata (scalability and reliability), our pipeline, and much of our data, are provided in accordance with open science.  The tested data originate from a wide assortment of institutions and projects, demonstrating the robustness of MIGRAINE.  We demonstrate the improved reliability of our pipeline over the previous state-of-the-art, in addition to its improved scalability.   Moreover, we utilize a notion of reliability related to the mean reciprocal rank often used in the information retrieval community. Finally, a useful metric for assessing pipeline and graph reliability is constructed and demonstrated.
\section{MIGRAINE Pipeline}

\begin{figure}[h!]
\centering
\includegraphics[width=8cm]{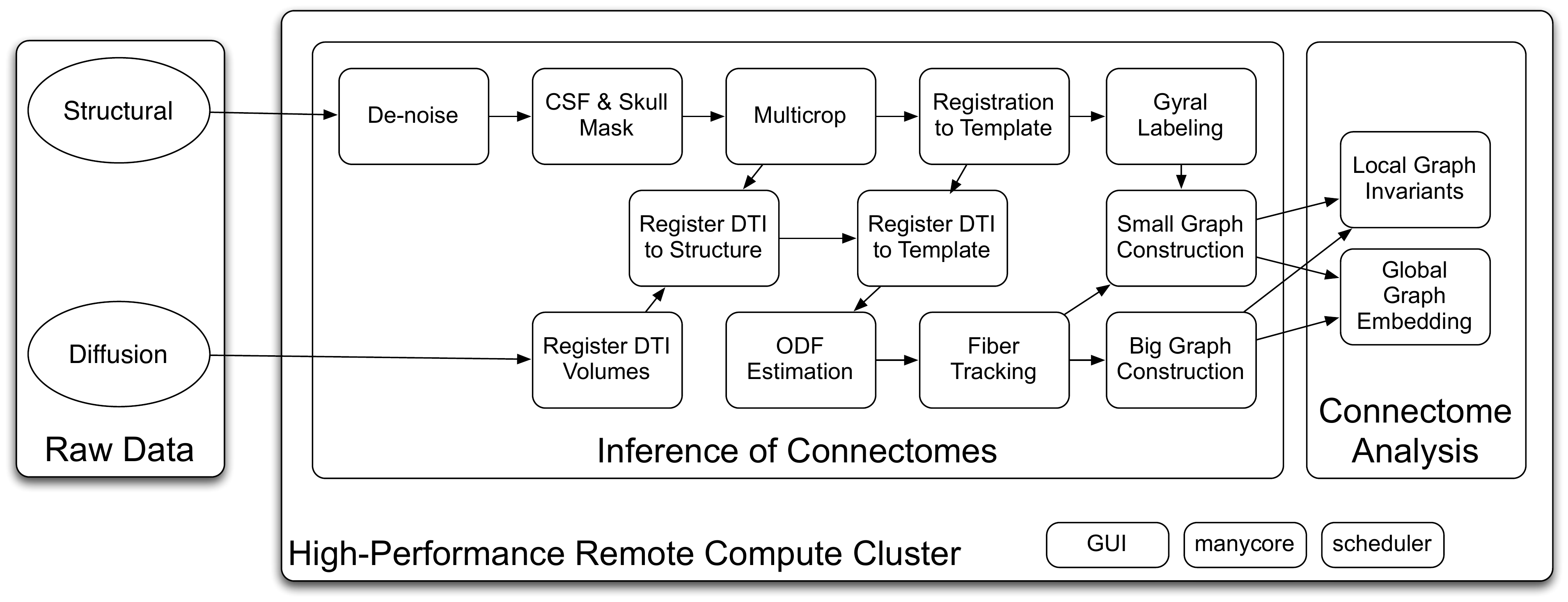}
\caption{MIGRAINE Pipeline Overview}
\label{fig:migrainefig}
\end{figure}

\subsection{Raw Data}

Table \ref{tab:data} provides summary statistics for the various datasets processed via MIGRAINE.  
For each dataset, we collected both diffusion and structural MRI (MPRAGE). The two test-retest datasets (KKI--42 and NKI-24) were used to assess reliability; the other datasets were processed because they contain interesting phenotypic information (covariates) that we will utilize in future studies.  
 
\begin{table}[h!]
	\caption{Various datasets successfully processed via MIGRAINE. Key for covariates: S=standard (sex, age, handedness), C=cognitive, B=behavioral, L=language, D=diagnostic (e.g., bipolar). The COINS column indicates whether the dataset will be made available via the Mind Research Network COINS data exchange (\url{http://coins.mrn.org/dx}). *=available at \url{www.nitrc.org/projects/multimodal}.  } \label{tab:data}
	\centering
\begin{tabular}{|c|c|c|c|c|}
	\hline
		\textbf{name} & \textbf{\# subjects} & \textbf{covariates} & \textbf{ref} & \textbf{COINS}   \\ \hline 
		KKI--42 & 21 & S & \cite{Landman2011} & * \\
		NKI--24 & 12 & S & \cite{milham2012} & Y \\
		MRN--111 & 111 & C & N/A & Y \\
		MRN--1313 & 1313 & C,D & N/A & N \\ 
		CASL--36 & 36 & C,B,L & N/A & N \\
		\hline
\end{tabular}
\end{table}

\subsection{Inference of Brain-Graphs}
\subsubsection{LONI Processing Framework}

The original Java Image Science Toolkit (JIST)-based \cite{Lucas2010} pipeline, consisting of 22 Java modules, is wrapped and integrated within the LONI pipeline framework \cite{Dinov2009}. Via swapping out some modules with improved functions, reducing I/O and communication, processing time is significantly reduced.
MIGRAINE is flexible and can be modified using existing neuroimaging modules already incorporated in LONI (e.g., \cite{Jenkinson2012}), or custom code that is command-line executable.  

\subsubsection{Small Graph Generation}

Our graph generation routines (summarized in Figure \ref{fig:migrainefig}) consist largely of the following steps.  Small graphs (e.g. 70 vertices and $\binom{70}{2}=2415$ edges) are computed as detailed in \cite{Gray2011}, except for additional steps to register the structural and diffusion data for each subject to a common template (e.g., the MNI atlas \cite{MNI2006}).  The voxels in the template brain volume are labeled in accordance with the Desikan atlas regions \cite{Desikan2006}.  We estimate tensors and perform deterministic tractography (FACT \cite{Mori1999}) to estimate fibers in the brain (note that both of these functions can easily be swapped out for more sophisticated, but time consuming, options).  Finally, an estimate of connectivity between each pair of regions is recorded (e.g.,  the number of times each region pair is connected by a fiber).  Much of the connectome literature uses \emph{region}-wise graphs, rather than \emph{voxel}-wise \cite{Craddock2013}.  These results therefore enable comparison with previous analysis, allow for the assessment of reliability between pipelines, and support existing classification methods.

\subsubsection{Big Graph Generation}

To generate big graphs, we utilize the Magnetic Resonance One-Click Pipeline (MROCP) code base as detailed in \cite{Mhembere}.  Initially a mask (e.g ROIs) is applied to the fiber streamlines created during small graph estimation. Each surviving voxel becomes a vertex in the sparse column compressed graph. Here edges represent a single fiber connecting a pair of vertices within the bounds defined by the mask. Finally, we iterate over each fiber streamline, recording an edge between every two vertices that can be reached (i.e. that are connected) by a single fiber. The entire codebase to generate the big graphs was developed and written in Python.\\ \\
All of these graphs are aligned, and for each MR scan we obtain a big graph with $\approx 10^7$ \emph{aligned} vertices and $\approx 10^{10}$ edges.  Because we are conservative with the mask, many of these voxels are noise.  We therefore reduce these graphs to their largest connected component, which keeps essentially all white matter voxels, consisting of $\approx 10^5$ vertices and $\approx 10^8$ edges. 

\subsection{Analysis of Brain-Graphs}

Computing analytics (i.e., multivariate glocal invariants \cite{Mhembere}) on big graphs is a challenging endeavor due to the computational intensity associated with processing graphs with $\approx 10^8$ edges.  Equivalent computational tasks are thus generally designated to specialized hardware like GPUs, graph processing engines like GraphLab \cite{Gonzalez}, or distributed solutions like MapReduce. We utilize MROCP to efficiently compute several multivariate graph analytics, including Latent Position (LP-k), Number of Local 3-Cliques (NL-3), Clustering Coefficient (CC), Scan Statistic-1 (SS-1), Degree and Edge count  (see \cite{Mhembere} for details). 
\subsection{Graph Reliability}

The literature on reliability focuses primarily on parametric reliability of scalar functions of the graphs.  In other words, they (implicitly) assume that the graphs themselves are reliable, and then ask questions about particular features of the graphs.  Because the data collection and graph generation processes are so noisy, we desire to assess the reliability of their composition.  Thus, given a fixed pipeline, we can compare reliability of two different scanners or scanning sequences.  Alternatively, given a fixed scanner and sequence, we can compare the reliability of two different pipelines.  

Let $\bxi \from \Omega \times \mathcal{T} \to \Xi$ be a brain-valued random variable.  In  words, $\omega \in \Omega$ denotes a particular person, and $t \in \Omega$ denotes a particular time, and $\bxi_t(\omega) \equiv \bxi(\omega,t)$ denotes person $\omega$'s \emph{actual} brain at time $t$.  Let $\psi \from \Xi \to \mathcal{X}$ be a particular \mmmr~ scanner and scanner sequence, so $x_t(\omega)=\psi(\bxi_t(\omega)) \in \mathcal{X}$ is the output from the scanner, and the input to MIGRAINE.  MIGRAINE converts $x$ to graphs, $\phi \from \mathcal{X} \to \mathcal{G}$.  Therefore, the scanning and pipeline together form the composition $f=\phi \,\circ\, \psi \from \Omega \times \mathcal{T} \to \mathcal{G}$, and we can test the reliability of either $\phi$ or $\psi$ over time for a particular subject, as follows.

Let $\delta \from \mc{G} \times \mc{G} \to \Real$ be a graph-value metric.  For vertex aligned graphs, we adopt the Frobenius norm of their adjacency matrices, $\delta(G,G')=\norm{A-A'}_F$, due to its simplicity and its theoretical properties (in particular, under that metric, a $k$-nearest neighbor classifier is universally consistent for graphs \cite{Vogelstein2009}).  Thus, given $n$ subjects, each with observations at two different times, we obtain $2n$ $x$'s as input to MIGRAINE, and we compute $\binom{2n}{2}$ distances (because distances are symmetric).  For each scan $i \in [2n]=\{1,\ldots, 2n\}$, we rank all remaining $2n-1$ scans using $\delta$. Let $i$ and $i'$ denote the first and second scan of subject $i$, respectively, and let $r_i \in [2n-1]$ denote the rank of $i'$ relative to $i$.  
We define reliability of $f$ as $R(f)= (2n - \sum_{i \in [2n]} r_i) / (2n-1) \in (0,1)$, so $R=1$ is maximally reliable and $R=0$ is minimally reliable.  Note that this notion of reliability is not limited to assessing $f$'s or graphs -- it is broadly applicable.  Moreover, it is  nonparametric and robust to outliers, and makes no distributional assumptions. 

\section{Results}

We successfully processed subjects from a variety of data corpora (both existing and new), totaling over 1500 subjects from multiple centers and acquisition paradigms. All of the resulting graphs and analytics are currently being used to develop classifiers, provide new insight into the way brains are wired and to determine which aspects of the network are informative in predicting cognitive properties. 

\subsection{Scalability}

\begin{figure}[h!]
\centering
\includegraphics[width=8cm]{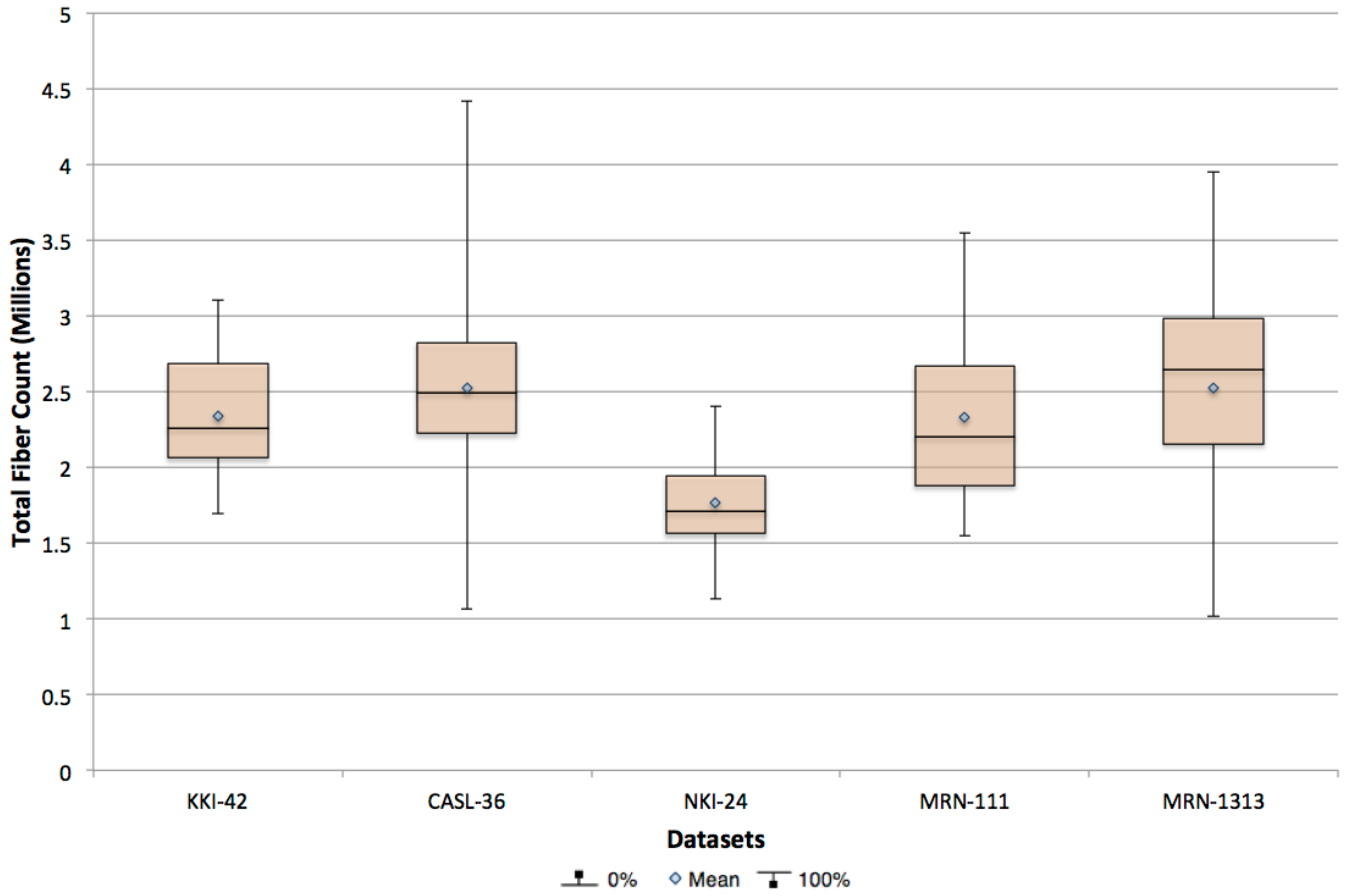}
\caption{Box plots for each data set, showing total fiber count for each dataset.}
\label{fig:boxplot}
\end{figure}

The current iteration of our software in the LONI Pipeline results in significant improvements to both scalability and processing time relative to the MRCAP baseline \cite{Gray2011}, which produces a small graph in approximately 10 hours on our small cluster (248 concurrent nodes, 1 TB total RAM).  On average, the MIGRAINE baseline takes approximately 3 hours/subject to compute small graphs (i.e., the output from MRCAP), an additional 5 hours/subject to produce big graphs, and 3.5 hours/subject for graph invariants, for a total of 11.5 hours/subject.  Much of this improvement is obtained by utilizing a common registration template, allowing for anatomical labels to be computed only once and then reused.  Multi-node capabilities only contribute marginally for a single subject (in both pipelines) because the most intensive computations occur serially.  However, there are significant efficiencies in scheduling when evaluating a large number of subjects, and the number of nodes is the limiting factor.  Run time for each of our datasets is presented in Table~\ref{tab:runtime}, including the results of a dataset with over 1000 subjects.  A univariate measure of total fiber count per subject is shown in Figure~\ref{fig:boxplot}.

\begin{table}[h!]
	\caption{Total and average run times for each dataset.} \label{tab:runtime}
  \begin{center}
  \resizebox{8.5cm}{!} {

  \begin{tabular}{l c|c|c|c|c|c|}
  \cline{3-7}
  & & \multicolumn{5}{ c| }{\textbf{Time (Hours)}} \\ \cline{3-7}
    \hline
 	\multicolumn{1}{|c|}{\textbf{}}&\multicolumn{1}{|c|}{\textbf{\#}}   & \textbf{Small} & \textbf{Big} &  &  & \textbf{Average/}\\ 
    \multicolumn{1}{ |l| }{\textbf{Dataset}} & \multicolumn{1}{|c|}{\textbf{scans}} & \textbf{Graphs} & \textbf{Graphs} & \textbf{Analytics} & \textbf{Total} & \textbf{Subject}\\ \hline
    \multicolumn{1}{ |l| }{\textbf{KKI-42}} & \multicolumn{1}{|c|}{\textbf{42}} & 2.9 & 4.7 & 3.9 & 11.5 & 11.2 \\
    \multicolumn{1}{ |l| }{\textbf{CASL-36}} & \multicolumn{1}{|c|}{\textbf{36}} & 3.1 & 4.6 & 3.7 & 11.3 & 11.0 \\
    \multicolumn{1}{ |l| }{\textbf{NKI-TRT}} & \multicolumn{1}{|c|}{\textbf{24}} & 3.2 & 5.2 & 3.4 & 11.8 & 11.6 \\
    \multicolumn{1}{ |l| }{\textbf{MRN-111}} & \multicolumn{1}{|c|}{\textbf{111}} & 1.5 & 6.1 & 3.0 & 10.5 & 9.8 \\
    \multicolumn{1}{ |l| }{\textbf{MRN-1313}} & \multicolumn{1}{|c|}{\textbf{1313}} & 9.9 & 37.3 & 18.3 & 65.5 & 10.2 \\
    \hline
  \end{tabular}
}
\end{center}
\end{table}

\subsection{Reliability}

A variety of tools have been developed to analyze intermediate pipeline products, such as matrix comparison tools and analysis of fiber counts.  MIGRAINE leverages several algorithmic improvements versus MRCAP, including changes to data preprocessing and the registration to a common registration space as described previously.  The results differed by 13\% from the MRCAP baseline and produced better subject separability as shown in Table~\ref{table:data_validation}.  
 
\begin{table}[h!]
	\caption{Validation showing improved discrimination relative to MRCAP using the KKI--42 dataset \cite{Landman2011}.} \label{table:data_validation}

 \begin{center}

\resizebox{8.5cm}{!} {
 \begin{tabular}{|>{\bfseries}c|c|c|c|c|}

    \hline
    & \textbf{Intra-Sub} & \textbf{Inter-Sub} & \textbf{Closest} & \textbf{\#} \\ 
    \textbf{Pipeline}   & \textbf{Diff} & \textbf{Mean Diff} & \textbf{Inter-Sub} & \textbf{Matches} \\\hline
     MRCAP & 26032 & 51584 & 38451 & 40/42 \\
    MIGRAINE & 20378 & 56126 & 42663 & 42/42 \\
    \hline
  \end{tabular}
  }
    \end{center}
\end{table}     
 
To validate that our graphs produce a repeatable signal, we used the KKI--42 Test-Retest Data \cite{Landman2011} to analyze graph estimation reliability, similar to \cite{Gray2011a}.  We demonstrated that the MIGRAINE pipeline produced a stable connectivity measurement across multiple scans of the same subject.

\begin{figure}[h!]
\centering
\includegraphics[width=7cm]{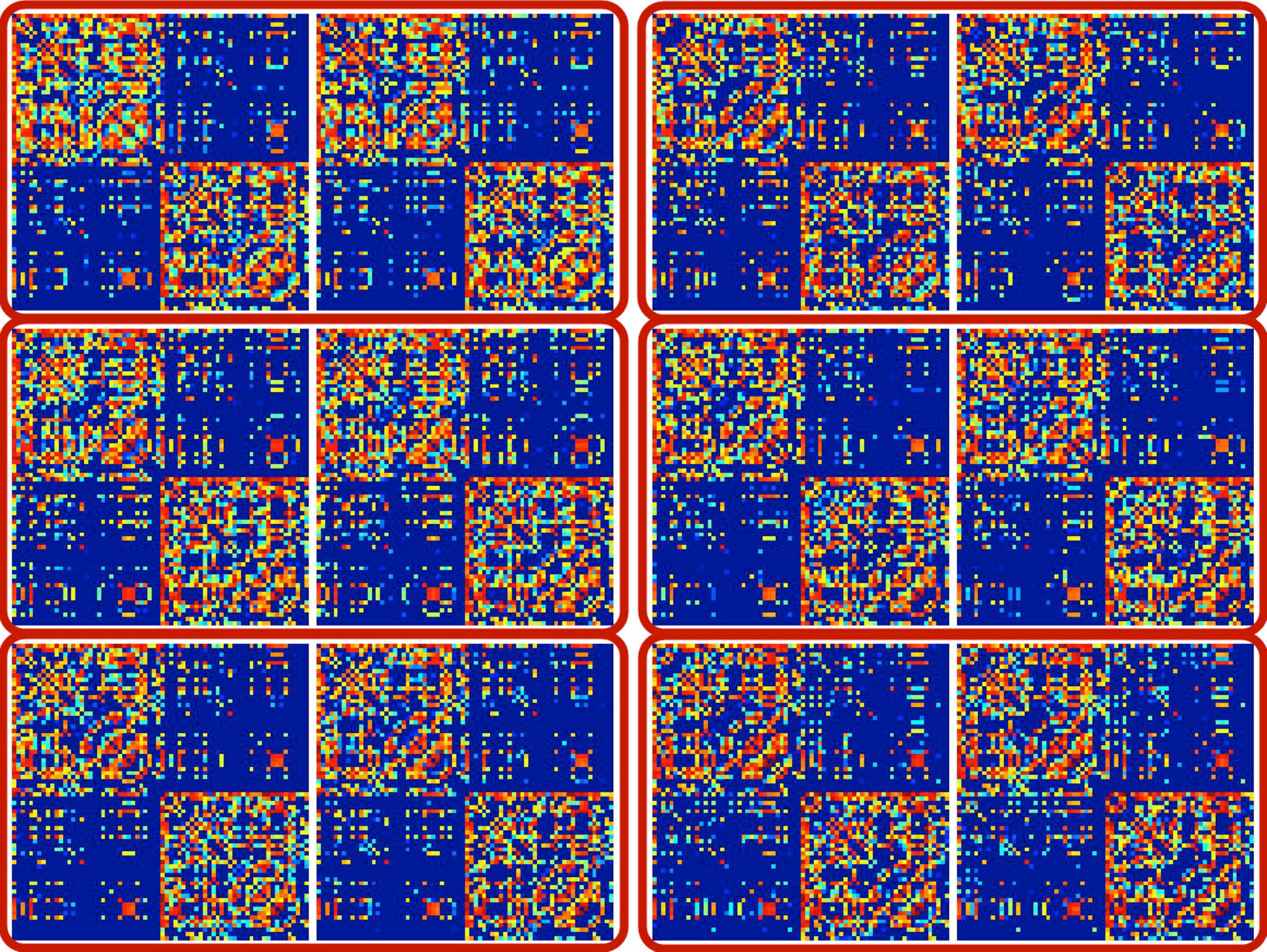}
\caption{Six Test-retest graphs. Top (L-R):  Male, 25 years old (M25), F26, Middle:  M25, F30, Bottom: M38, F61.}
\label{fig:trtimage}
\end{figure}

\begin{figure}[h!]
\centering
\includegraphics[width=7cm]{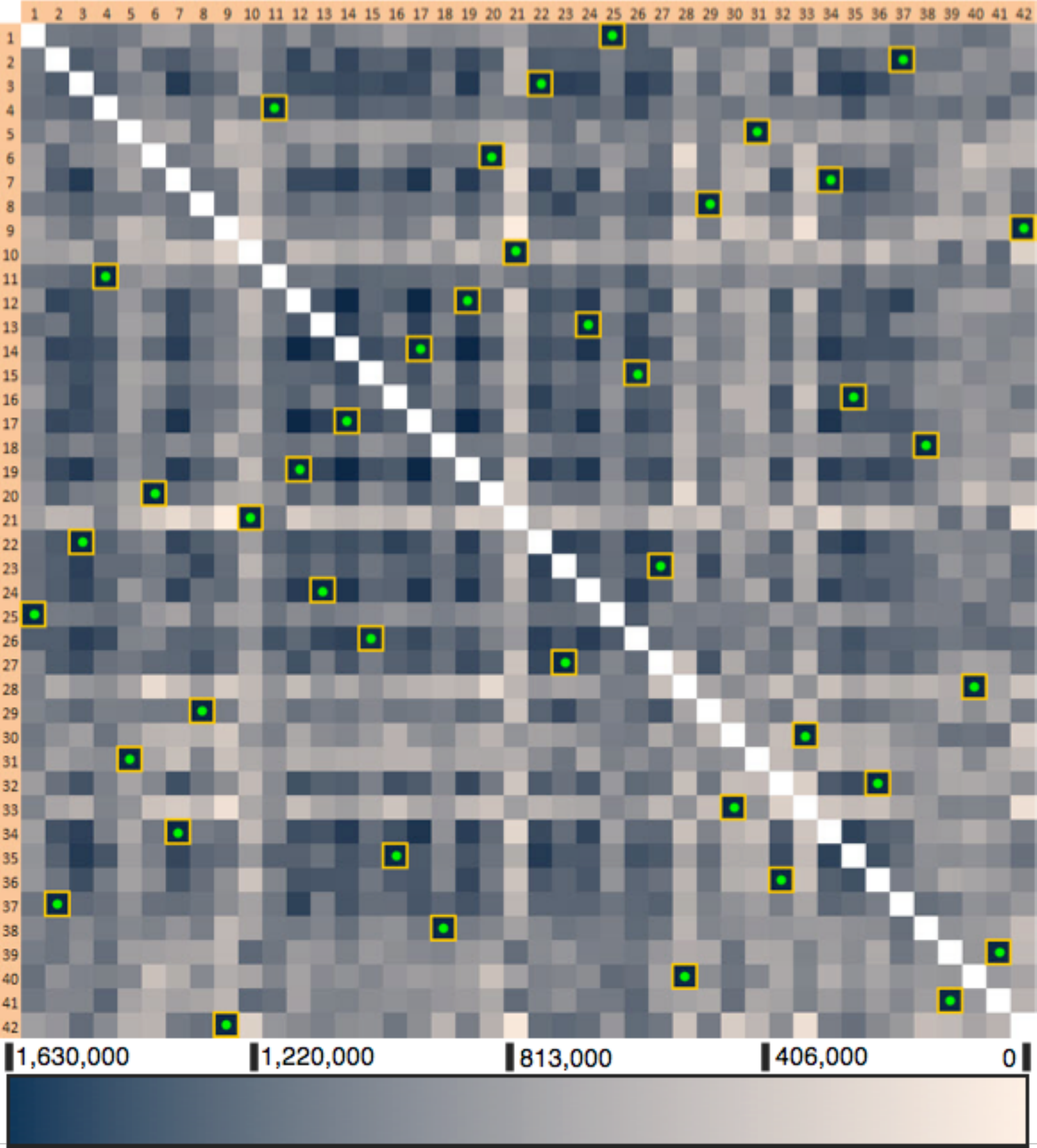}
\caption{KKI Test-Retest Results.  Yellow boxes: Highest similarity, Green dots: True pairs, White: Self-comparison.}
\label{fig:kki21vis}

\end{figure}

For all 42 graphs, the most closely related graph (as computed with the Frobenius norm) belonged to the same person, scanned at a different time.  A visualization of the graphs for six test-retest pairs are shown in Figure~\ref{fig:trtimage}, and the results of all individual subject comparisons are shown in Figure~\ref{fig:kki21vis}.  

\section{Conclusions}

MIGRAINE is robust and has been utilized to process over 1500 subjects from a variety of datasets in a rapid, extensible, automated framework. In addition to producing small graphs, we have demonstrated additional processing capability through the estimation of big graphs and analytics.  The pipeline is scalable and has internal validation and packaging scripts to enable efficient analysis.  Finally, we provide evidence of classification signal in the estimated brain graphs. 
\section*{Acknowledgments}

{\small
The authors thank the Image Analysis and Communications Laboratory at The Johns Hopkins University, Baltimore, MD. 
This work has been supported by the NSA Research Program on Applied Neuroscience and NIH/NINDS 5R01NS056307. 
John Templeton Foundation entitled ``The Neuroscience of Creativity'' funded the MRN--111 dataset collection.
}

\bibliographystyle{IEEEtran}	
\bibliography{dolphinieee.bib}		

\begin{thebibliography}{10}
\providecommand{\url}[1]{#1}
\csname url@samestyle\endcsname
\providecommand{\newblock}{\relax}
\providecommand{\bibinfo}[2]{#2}
\providecommand{\BIBentrySTDinterwordspacing}{\spaceskip=0pt\relax}
\providecommand{\BIBentryALTinterwordstretchfactor}{4}
\providecommand{\BIBentryALTinterwordspacing}{\spaceskip=\fontdimen2\font plus
\BIBentryALTinterwordstretchfactor\fontdimen3\font minus
  \fontdimen4\font\relax}
\providecommand{\BIBforeignlanguage}[2]{{%
\expandafter\ifx\csname l@#1\endcsname\relax
\typeout{** WARNING: IEEEtran.bst: No hyphenation pattern has been}%
\typeout{** loaded for the language `#1'. Using the pattern for}%
\typeout{** the default language instead.}%
\else
\language=\csname l@#1\endcsname
\fi
#2}}
\providecommand{\BIBdecl}{\relax}
\BIBdecl

\bibitem{Sporns2010}
\BIBentryALTinterwordspacing
O.~Sporns, ``{Networks of the Brain},'' \emph{Learning}, no. August, p. 375,
  2010.
\BIBentrySTDinterwordspacing

\bibitem{Lichtman2008}
\BIBentryALTinterwordspacing
J.~W. Lichtman and J.~R. Sanes, ``{Ome sweet ome: what can the genome tell us
  about the connectome?}'' \emph{Current opinion in neurobiology}, vol.~18,
  no.~3, Jun. 2008.
\BIBentrySTDinterwordspacing

\bibitem{VanEssen2012a}
\BIBentryALTinterwordspacing
D.~C. {Van Essen} \emph{et~al.}, ``{The Human Connectome Project: a data
  acquisition perspective.}'' \emph{NeuroImage}, vol.~62, no.~4, pp. 2222--31,
  Oct. 2012.
\BIBentrySTDinterwordspacing

\bibitem{Mennes2012}
\BIBentryALTinterwordspacing
M.~Mennes, B.~B. Biswal, F.~X. Castellanos, and M.~P. Milham, ``{Making data
  sharing work: The FCP/INDI experience.}'' \emph{NeuroImage}, Oct. 2012.
\BIBentrySTDinterwordspacing

\bibitem{Gray2011}
W.~R. Gray \emph{et~al.}, ``{Magnetic Resonance Connectome Automated
  Pipeline},'' \emph{Pulse}, no. APRIL, pp. 1--5, 2012.

\bibitem{Cui2013}
\BIBentryALTinterwordspacing
Z.~Cui \emph{et~al.}, ``{PANDA: a pipeline toolbox for analyzing brain
  diffusion images.}'' \emph{Frontiers in human neuroscience}, vol.~7, no.
  February, p.~42, Jan. 2013.
\BIBentrySTDinterwordspacing

\bibitem{Braun2012}
U.~Braun \emph{et~al.}, ``{Test-retest reliability of resting-state
  connectivity network characteristics using fMRI and graph theoretical
  measures.}'' \emph{Neuroimage}, vol.~59, no.~2, pp. 1404--1412, 2012.

\bibitem{Landman2011}
\BIBentryALTinterwordspacing
B.~A. Landman \emph{et~al.}, ``{Multi-parametric neuroimaging reproducibility:
  a 3-T resource study.}'' \emph{NeuroImage}, vol.~54, no.~4, pp. 2854--66,
  Mar. 2011.
\BIBentrySTDinterwordspacing

\bibitem{milham2012}
\BIBentryALTinterwordspacing
K.~B. Nooner \emph{et~al.}, ``{The NKI-Rockland Sample: A Model for
  Accelerating the Pace of Discovery Science in Psychiatry},'' \emph{Frontiers
  in Neuroscience}, vol.~6, no. 152, 2012.
\BIBentrySTDinterwordspacing

\bibitem{Lucas2010}
\BIBentryALTinterwordspacing
B.~C. Lucas \emph{et~al.}, ``{The Java Image Science Toolkit (JIST) for rapid
  prototyping and publishing of neuroimaging software.}''
  \emph{Neuroinformatics}, vol.~8, no.~1, pp. 5--17, Mar. 2010.
\BIBentrySTDinterwordspacing

\bibitem{Dinov2009}
\BIBentryALTinterwordspacing
I.~D. Dinov \emph{et~al.}, ``{Efficient, Distributed and Interactive
  Neuroimaging Data Analysis Using the LONI Pipeline.}'' \emph{Frontiers in
  neuroinformatics}, vol.~3, p.~22, Jan. 2009.
\BIBentrySTDinterwordspacing

\bibitem{Jenkinson2012}
\BIBentryALTinterwordspacing
M.~Jenkinson \emph{et~al.}, ``{FSL},'' \emph{NeuroImage}, vol.~62, no.~2, pp.
  782--790, 2012.
\BIBentrySTDinterwordspacing

\bibitem{MNI2006}
G.~Grabner \emph{et~al.}, ``{Symmetric Atlasing and Model Based Segmentation:
  An Application to the Hippocampus in Older Adults},'' in \emph{Lecture Notes
  in Computer Science}, 2006, vol. 4191, pp. 58--66.

\bibitem{Desikan2006}
\BIBentryALTinterwordspacing
R.~S. Desikan \emph{et~al.}, ``{An automated labeling system for subdividing
  the human cerebral cortex on MRI scans into gyral based regions of
  interest.}'' \emph{NeuroImage}, vol.~31, no.~3, pp. 968--80, Jul. 2006.
\BIBentrySTDinterwordspacing

\bibitem{Mori1999}
\BIBentryALTinterwordspacing
S.~Mori, B.~Crain, V.~Chacko, and P.~{Van Zijl}, ``{Three-dimensional tracking
  of axonal projections in the brain by magnetic resonance imaging},''
  \emph{Annals of neurology}, vol.~45, no.~2, pp. 265--269, Feb. 1999.
\BIBentrySTDinterwordspacing

\bibitem{Craddock2013}
\BIBentryALTinterwordspacing
R.~C. Craddock \emph{et~al.}, ``{Imaging human connectomes at the
  macroscale.}'' \emph{Nature methods}, vol.~10, no.~6, pp. 524--39, Jun. 2013.
\BIBentrySTDinterwordspacing

\bibitem{Mhembere}
D.~Mhembere \emph{et~al.}, ``{Computing Scalable Multivariate Glocal Invariants
  of Large (Brain-) Graphs},'' \emph{IEEE GlobalSIP, Accepted}, 2013.

\bibitem{Gonzalez}
J.~E. Gonzalez, D.~Bickson, and C.~Guestrin, ``{PowerGraph : Distributed
  Graph-Parallel Computation on Natural Graphs},'' pp. 17--30, 2012.

\bibitem{Vogelstein2009}
\BIBentryALTinterwordspacing
J.~T. Vogelstein, R.~J. Vogelstein, and C.~E. Priebe, ``{Are mental properties
  supervenient on brain properties?}'' \emph{Nature Scientific Reports}, p.~11,
  2011.
\BIBentrySTDinterwordspacing

\bibitem{Gray2011a}
W.~Gray \emph{et~al.}, ``{Magnetic Resonance Connectome Automated Pipeline and
  Repeatability Analysis},'' \emph{Society for Neuroscience Abstract}, 2011.

\end{thebibliography}

\end{document}